# Determination of the crystal field splitting energy in Cd$_3$As$_2$ using magnetooptics

G. Krizman[1], T. Schumann[2], S. Tchoumakov[3], B.A. Assaf[1,4], S. Stemmer[2], L.A. de Vaulchier[1], Y. Guldner[1]

[1] Laboratoire de Physique de l'Ecole normale supérieure, ENS, Université PSL, CNRS, Sorbonne Université, 24 rue Lhomond 75005 Paris, France

[2] Materials Department, University of California, Santa Barbara, CA 93106, USA

[3] Département de Physique, Université de Montréal, Quebec, H3C 3J7, Canada

[4] Department of Physics, University of Notre Dame, Notre Dame, IN 46556, USA

**Symmetry considerations are of extreme importance to the topological properties of crystals. A crystal field splitting $\delta$ yields Dirac nodes near the Brillouin zone center in Cd$_3$As$_2$, but its value has yet to be determined with precision. We study the band structure of Cd$_3$As$_2$ using magnetooptical infrared spectroscopy measurements on epilayers with low carrier density grown by molecular beam epitaxy. By combining angular dependent cyclotron resonance with Landau level spectroscopy measurements in the Faraday geometry, we determine that $\delta$ is positive and equal to $15 \pm 5$ meV in Cd$_3$As$_2$. Our results lead to a more accurate knowledge of the details of the band structure of this Dirac semimetal such as the position its Dirac nodes in momentum space and their splitting into Weyl nodes under a magnetic field.**

It is well established that Cd$_3$As$_2$ is a three-dimensional semimetal. [1–12] Experimental measurements on this material have unambiguously revealed the presence of electrons with ultra-low effective mass ($< 0.03 m_0$) [3,13] and record high mobility ($\sim 10^5$ cm²/Vs) in single crystals. [1] A magnetic field was argued to yield Weyl fermions in Cd$_3$As$_2$. [14,15] As a result, a large number of transport studies have been performed on Cd$_3$As$_2$ at high magnetic fields to probe the shape of the Fermi surface, the Fermi arcs, and other possible signatures of Weyl fermions. [16–23] As Cd$_3$As$_2$ is generally n-doped due to native defects. [24] [7] Achieving charge neutrality in the bulk system has proven difficult. The details of the band dispersion in the vicinity of Dirac nodes thus remained poorly understood. Such details of the band dispersion are essential to understand and control the chiral behavior of any possible Weyl fermions arising in Cd$_3$As$_2$.

Several theoretical models consistently consider the band dispersion of Cd$_3$As$_2$ to be dominated by the s-p interaction of energy bands in the vicinity of the Γ-point, as in the case of III-V semiconductors. [6,25–28] However, contrary to the III-V family, which is cubic, Cd$_3$As$_2$ has a tetragonal crystal structure. This tetragonal lattice distortion introduces a crystal field splitting that raises the heavy hole band above the conduction band at the Γ-point yielding two stable band crossings at $\pm k_0$ along the $k_z$ direction. [12,26,28] This description is qualitatively identical in many of the theoretical works that have been so far published on Cd$_3$As$_2$. Quantitatively, the exact value of the band parameters, most notably the s-p gap $E_g$, the crystal field splitting $\delta$, the $\boldsymbol{k} \cdot \boldsymbol{p}$ matrix element along the tetragonal axis and its anisotropy differ in different theoretical studies and experimental measurements are required to constrain them.

Previous magnetooptical measurements have attempted to independently measure these four parameters and have succeeded in providing a reliable upper bound for $\delta$ given $E_g$ and the matrix element. [6,29] It is challenging to accurately determine the four parameters independently since the Fermi energy in Cd$_3$As$_2$ is generally on the order of $E_g$ and likely larger than $\delta$. [29] Also, while previous angular resolved photoemission and scanning tunneling microscopy experiments have measured the Dirac velocity of $Cd_3As_2$, [3,8] the surface sensitive character of these methods, and their lack of precision close to the 10 meV range, makes it difficult to employ them to study the 3D dispersion of $Cd_3As_2$.

In this work, we build on the improved quality obtained from growing epilayers of (112) oriented $Cd_3As_2$ by molecular beam epitaxy (MBE) on GaSb/GaAs (111) substrates to study the 3D band structure of $Cd_3As_2$ near the Γ-point. [30] Using a combination of magnetooptical infrared spectroscopy with $\boldsymbol{B} \perp$ (112) and angular dependent cyclotron resonance measurements, we are able to accurately extract $E_g$, $\delta$, the matrix element in the (001) plane $P_\perp$ and the anisotropy of the matrix element $\eta^2$ (defined later on). The determination of the four parameters is of major importance to further study the impact of magnetic field on the band dispersion of $Cd_3As_2$ and its role in promoting or suppressing Weyl fermions.

**Results**

**Magnetooptical spectroscopy with perpendicular magnetic field**

Magnetooptical infrared spectroscopy measurements are performed up to 15 T and for energies up to 500 meV using the setup described in our previous work. [31] Spectra are taken at 4.2 K and in the Faraday geometry. In this configuration, conventional dipole selection rules yield transitions between Landau levels of identical spin and indices different by 1. We first focus on measurements taken with the magnetic field along the [221]-direction, which corresponds to the normal of the (112) plane in a tetragonal structure. Results for two epilayers (280 nm and 450 nm thick) are shown in Fig. 1(a,b) respectively. Several transmission minima that shift to higher energy with increasing field are observed. They correspond to optical transitions between Landau levels. Spectra consist of two dominant absorptions marked by the blue and red arrows and several transmission minima of smaller amplitude.

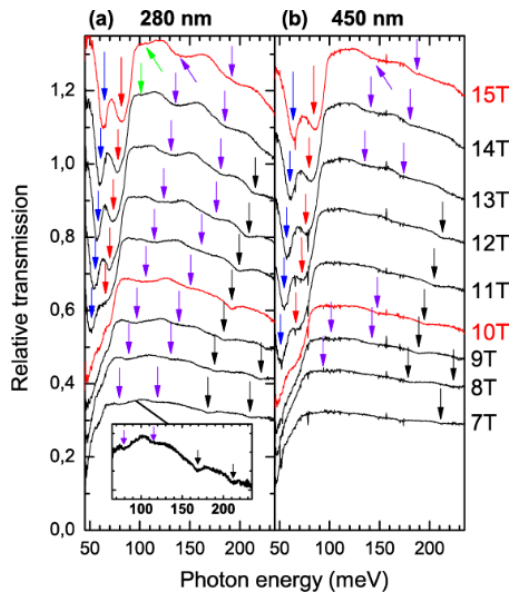

**FIG 1.** Magnetooptical spectra in the mid-infrared for magnetic fields between 7 T and 15 T measured in a 280 nm **(a)** and a 450 nm **(b)** $Cd_3As_2$ epilayer. Transmission minima are marked by colored arrows: blue and red denote two cyclotron resonance transitions, purple and black denote two series of interband transitions, and green stands for the combined resonance. The inset in (a) is a vertical zoom (x3) of the 7 T spectrum.

Figure 2(a,b) shows the transition fan charts obtained by pin-pointing the energy position of the minima seen in Fig. 1(a,b) (dots). The blue and red transitions behave like cyclotron resonance lines of non-parabolic materials and exhibit a non-linear variation with magnetic field. In addition, the purple and black series likely extrapolate to two different energies at the zero magnetic field evidencing the presence of at least three energy bands contributing optical transitions near the band edges. To further analyze the Landau levels of $Cd_3As_2$, a precise description of its band structure near the Γ-point is required.

### $k.p$ model for Landau levels

$Cd_3As_2$ is a tetragonal material with a rotational invariance along [001], parallel to the $z$ axis in the following. The $Cd_3As_2$ band structure has been obtained by DFT calculations [9,12,32] or by using a modified 8-band Kane model. [6,25,26] To compute the Landau levels, a low energy $k.p$ model is required. As detailed in Appendix A, the Kane Hamiltonian for cubic symmetry can be used to describe the Landau levels of tetragonal $Cd_3As_2$ by adding a parameter $\delta$ that represents the crystal field splitting of the p-bands. [25,26] In the Kane model, an s-like band is typically separated from a degenerate p-like band by an energy gap $E_g$. In $Cd_3As_2$, $E_g < 0$, meaning that the s-level is a valence band. Without the crystal field splitting, the band structure resembles that of semimetallic HgTe, [33] with two p-bands touching at the Γ-point. The p-levels are spin-orbit split with an energy $\Delta_{SO}$ fixed to 400 meV in this work. One of those is a light $p^{\pm 1/2}$ level that constitutes the conduction band and the second is a heavy (quasi-flat) $p^{\pm 3/2}$ level that yields a valence band. The value and sign of $\delta$ shift the $p^{\pm 3/2}$ valence band with respect to the $p^{\pm 1/2}$ conduction band. If $\delta > 0$, the conduction and valence band cross at two k-points yielding two Dirac cones. In this model, in addition to $E_g$ and $\delta$, the band dispersion is mainly governed by the Kane matrix elements $P_\parallel$ in the [001] direction and $P_\perp$ in the (001) plane. They are related by the anisotropy factor $\eta^2 = P_\perp/P_\parallel$.

The band structure ($E$ vs $k_z$) resulting from this model is shown in Fig. 2(c). Qualitatively, this model agrees with previous DFT and $k.p$ treatments at low energy in the vicinity of the Γ-point. In particular, the origin of the two Dirac nodes in $Cd_3As_2$ in this model is similar to that discussed by Wang *et al.* and Cano *et al.* in their theoretical work where a 4-band $k.p$ Hamiltonian with far-bands is used. [12,28] The main limitation of our description is the neglected role of the far-bands that yield the finite curvature of the heavy-hole band. We thus follow the convention of ref. [6] and set the heavy-hole band to be flat. Experimentally, this only implies that heavy-hole Landau levels are flat versus field, which approximately holds up to our maximum field (15 T). The remote-band effects can be introduced in our model as described in Appendix B. These effects essentially consist in taking into account perturbation from bands located at higher energy (typically ~1 eV in narrow-gap semiconductors or semimetals).

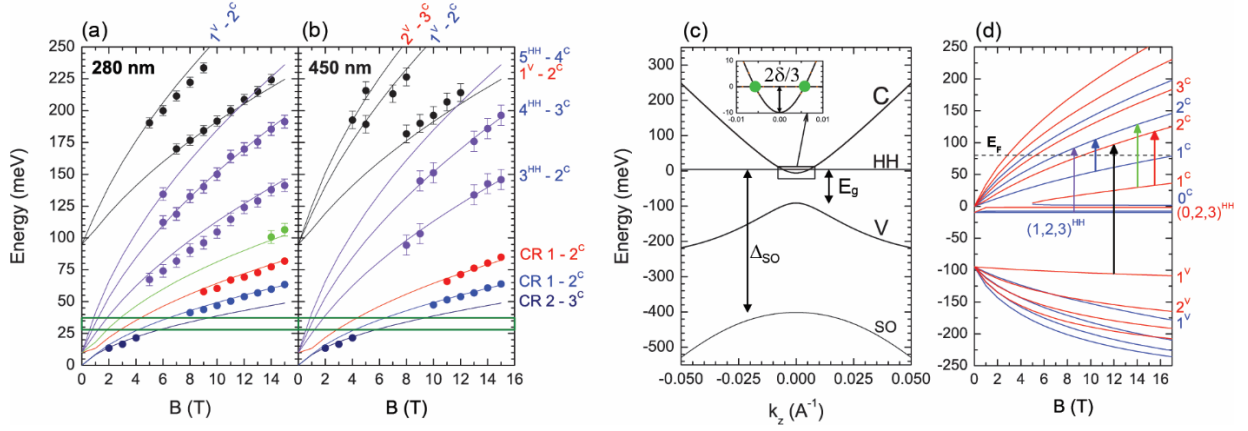

**FIG 2.** Magnetooptical fan chart for **(a)** a 280 nm and **(b)** a 450 nm Cd₃As₂ epilayer at 4.2 K. Error bars correspond to the full width at half maximum of the absorptions. **(c)** Energy dispersion ($E$ vs $k_z$) computed using the parameters obtained from the experimental fit. A zoom-in is shown in the insert with the two Dirac points in green. **(d)** Landau level energy at $q_B = 0$ versus magnetic field along [221]. Red and blue curves represent the different Zeeman components. Blue and red arrows denote CR $1^c$-$2^c$ for opposite spin quantum number. The lowest observed interband transitions for V-C and HH-C are shown as black and violet arrows respectively. The combined resonance is show as a green arrow.

Our samples are (112)-oriented. In this orientation the angle between the surface normal (the [221]-direction) and [001] is $\theta_0 = 54.7°$. Therefore, a perpendicular magnetic field makes an angle $\theta_z = \theta_0$ with respect to [001]. To interpret our data shown in Fig. 1, we need to calculate the Landau levels for a magnetic field tilted away from $k_z$. Based on the theoretical work of Wallace [27] and taking into account the anisotropy effect, the Landau levels are obtained by solving the following secular equation:

$$\gamma(E) = \frac{\text{sign}\left(f_{1\,\theta_z}(E)\right) eB(2n+1)}{\hbar} \sqrt{f_{1\,\theta_z}(E)[\cos(\alpha)^2 f_{1\,\theta_z}(E) + \sin(\alpha)^2 f_{2\,\theta_z}(E)]} + \frac{f_{1\,\theta_z}(E) f_{2\,\theta_z}(E)}{\cos(\alpha)^2 f_{1\,\theta_z}(E) + \sin(\alpha)^2 f_{2\,\theta_z}(E)} q_B^2$$

$$\pm \frac{P_\perp eB\Delta_{SO}}{3\hbar} \sqrt{P_\perp^2 \cos(\theta_z)^2 (E+\delta)^2 + E^2 P_\parallel^2 \sin(\theta_z)^2} \quad \textbf{(1)}$$

where we have introduced the following notations:

$$\alpha = \arctan[\eta^2 \tan\theta_z]$$

$$P = \sqrt{|P_\perp^2 \cos(\alpha)\cos(\theta_z) + P_\parallel P_\perp \sin(\alpha)\sin(\theta_z)|}$$

$$\gamma(E) = E(E - E_g)[E(E + \Delta_{SO}) + \delta(E + 2\Delta_{SO}/3)]$$

$$f_{1\theta_z}(E) = P^2[E(E + 2\Delta_{SO}/3) + \delta(E + \Delta_{SO}/3)]$$

$$f_{2\,\theta_z}(E) = P^2 E(E + 2\Delta_{SO}/3)$$

Here, $n$ is the Landau level index, and $q_B$ is the wave vector along the magnetic field. Eq. (1) shows a Zeeman-like band splitting (the term in front of $\pm$). By numerically solving Eq. (1) for each $n$ value, we obtain the Landau level energies. Fig. 2(d) shows the computed Landau levels at $q_B = 0$ for $\boldsymbol{B}//[221]$. Two types of transitions are of interest to this work: the interband transitions in violet and black, and the intraband transitions (or cyclotron resonances) in red and blue. The combined resonance (CombR) shown in green is discussed later. The calculated transition energies are shown by solid lines in Figs. 2(a,b). Good agreement is found for $|E_g| = 95 \pm 5$ meV, $P_\perp = 7.96 \pm 0.25$ eV.Å, $|\delta| = 15 \pm 15$ meV and $\eta^2 = 1.20 \pm 0.05$. A Fermi energy of 70 meV for the 280 nm sample and 80 meV for the 450 nm sample can also be deduced by looking at the field where the $2^c$ level in Fig. 2(d) crosses the Fermi energy and activate the red CR, shown in Figs. 2(a,b). Indeed the cyclotron resonance can only be observed when the final level is unoccupied i.e. above 9T in the 280nm sample and above 11T in the 450nm sample. In Fig. 2(d), notice that the $0^c$ and $1^c$ levels start only at 5 T. This is an artefact of our model with a flat heavy-hole band that yields an open Fermi surface close to the band crossing. This prevents quantization for certain directions of magnetic field and can be avoided by accounting for a finite heavy-hole mass (Appendix B). We do not observe any transitions involving those levels near 5T, therefore, this issue does not influence our analysis.

In Fig. 2(d), the 0c and 1c levels start only at 5 T. This is an artefact of our model with a flat heavy-hole band, which yields an open Fermi surface close to the band crossing which prevents quantization for certain directions of magnetic field. This could be avoided by accounting for a finite heavy-hole mass as described in Appendix B. We do not observe any transitions involving those levels near 5 T, therefore, this issue does not influence our analysis.Our determination of $E_g$ and $P_\perp$ are complementary to the recent work on $Cd_3As_2$. The precision in the determination of both parameters is a result of improved sample quality. To get a comparative view between our parameters and those determined from magnetotransport measurements, we extract the Kane velocity from Kane matrix element:

$$v_\perp = \sqrt{\frac{2}{3}\frac{P_\perp}{\hbar}} = 9.9 \times 10^5 \text{m/s}$$

The high velocity of the conduction band carriers in $Cd_3As_2$ and the small value of $|E_g|$ are necessary requirements to obtain a low carrier cyclotron mass. The latter is needed to accurately fit the two dominant CR transitions we observed (see Fig. 2(a,b)). Our determination of $v_\perp$ agrees with previous magnetooptical, [6] angle resolved photoemission and scanning tunneling microscopy measurements. [3] The anisotropy factor also agrees with previous Shubnikov-de-Haas measurements. [13] The value of $\delta$ determined from the fit to the fan-chart lacks precision. Any value between $-30$ meV and 30 meV would agree with our data. If $\delta < 0$, the bands do not intersect and the Dirac nodes predicted for $Cd_3As_2$ cannot exist. [12]

**Angular dependent cyclotron resonance**

To improve our determination of $\delta$, we carried out additional measurements of the angular dependence of the cyclotron resonance lines. As represented in Fig. 3(a), tilting **B** by an angle $\theta = 40$ ° with respect to the normal of the sample surface, we can vary $\theta_z$ (the angle between [001] and **B**) by rotating the sample about its surface normal by an azimuthal angle $\varphi$. We can thus check the anisotropy of the CR masses. In this geometry, $\theta_z$ is related to $\varphi$ by the following expression:

$$\cos(\theta_z) = \cos(\theta_0)\cos(\theta) - \sin(\varphi)\sin(\theta_0)\sin(\theta) \qquad (2)$$

$\theta_0$ is the angle between [221] and [001]. It can be shown (Appendix C) that $\delta$ impacts the shape of the Fermi surface and increases its anisotropy along the $k_z$ momentum direction. For $\delta \geq 0$, the position of the Dirac points along the Z-Γ-Z direction $\pm k_0$ is well approximated by: [20]

$$k_0 = \sqrt{|E_g|\delta/P_\parallel^2}$$

Therefore, if $\delta$ is increased, $k_0$ increases and the two Dirac fermion valleys move further apart. $\delta$ also impacts the shape of the Fermi surface and increases its anisotropy along the $k_z$ momentum direction. Figure 3(b) shows the anisotropy of the Fermi surface computed for $\delta = 0$ and $\delta = 15$ meV for a fixed $\eta^2 = 1.20$. In addition to the matrix element anisotropy factor $\eta^2$, $\delta$ yields an additional source of anisotropy. This is a consequence of $\delta$ altering the cyclotron mass along the tetragonal direction. We can measure this effect by probing the anisotropy of the cyclotron resonance observed in our experiment. The fact that we are able to resolve two different CR transitions occurring at different energies allows us to disentangle the effect of $\delta$ and $\eta^2$ on the anisotropy using angular dependent cyclotron resonance (CR) measurements.

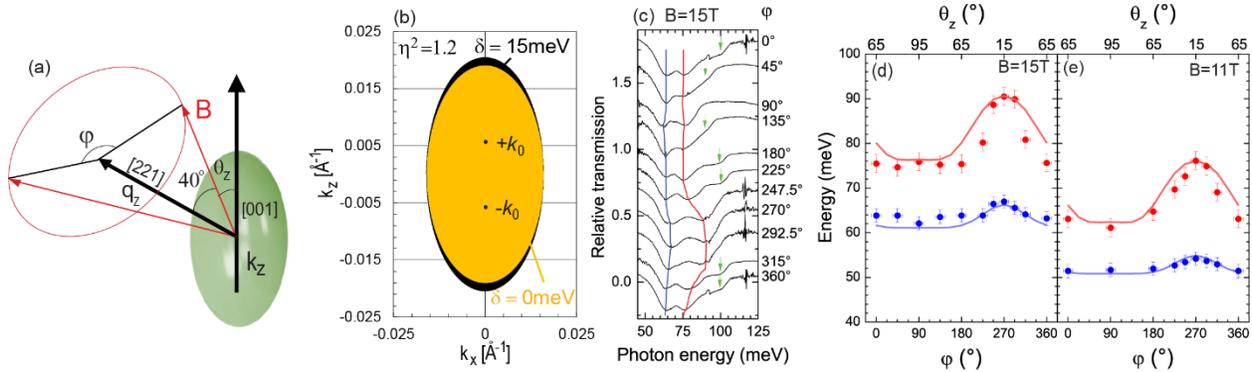

**FIG 3. (a)** Ellipsoidal Fermi surface illustrating the measurement geometry used for the angular dependence. **(b)** Projected Fermi surface computed for $\delta = 0$ meV (orange) and $\delta = 15$ meV (black) for $\eta^2 = 1.20$ and $E_F = 70$ meV. **(c)** Magnetooptical spectra taken at different $\varphi$ at $B = 15$ T and $\theta = 40$ °. The blue and red guide-for-the-eyes show the variation of the cyclotron energy (1-2$^c$(blue) and CR 1-2$^c$(red)) versus $\varphi$, which is also represented by the dots in **(d)** for $B = 15$ T and in **(e)** for $B = 11$ T. The red and blue solid lines in **(d)** and **(e)** are the calculated azimuthal variation of the CR energies.

Figure 3(c) shows magnetooptical spectra taken at 15 T at different $\varphi$ angles for fixed $\theta = 40$ °. The blue and red solid lines mark the transmission minima associated with CR 1-2$^c$(blue) and CR 1-2$^c$(red). Their energy variation as a function of $\varphi$ is shown by the dots in Fig. 3(d,e). A peak is observed at $\varphi = 270$ ° for both transitions corresponding to an angle $\theta_z = 15$ ° following Eq. (2). A minimal cyclotron mass is thus obtained for **B** aligned closest to [001]. To quantify this angular variation, we solve the secular equation (Eq. (1)) for different angles $\theta_z$. The resulting Landau levels are then used to compute the CR energies that are compared to the data. The computed angular dependent CR energies are compared to the data in Fig. 3(d,e) at $B = 15$ T and 11 T respectively. Using the two cyclotron resonances observed in the experiment, $\eta^2$ and $\delta$ are determined independently. The calculated maximum of the peak observed in Fig. 3(d,e) is found to depend significantly on $\delta$ and the solid lines in Fig. 3(d,e) show the best fit obtained with $\eta^2 = 1.20 \pm 0.02$ and $\delta = 15 \pm 5$ meV. To illustrate the strong $\delta$-dependence of the fit, Figure 4 shows the calculated angular dependence of the CR 1-2$^c$(red) for various $\delta$ magnitude and sign at $B = 11$ T.

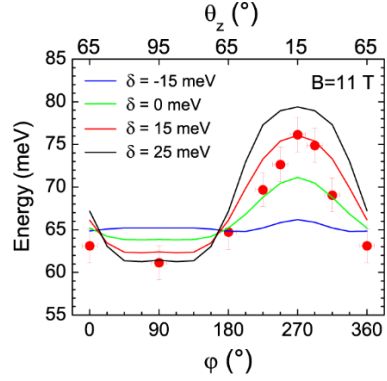

**FIG 4.** Calculated angular dependence of the CR 1-2$^c$(red) at $B = 11$ T for different magnitude and sign for $\delta$ (solid lines). The best agreement with experimental data (in red dots) is obtained for $\delta = +15$ meV.

Lastly, an additional absorption line marked by green arrows is observed in Fig. 3(c). It vanishes for specific $\varphi$ values. This absorption is attributed to the combined resonance CombR (the intraband transition 1$^c$(red)-2$^c$(blue)), also shown in green in Fig. 2(d). CombR is allowed due to the existence of an inter-spin coupling term appearing in the off-diagonal of the Hamiltonian. These terms are proportional to $\sin[2\arctan(\eta^2 \tan\theta_z)]$, as shown in Appendix D. The inter-spin coupling is hence suppressed for $\theta_z = 0$ ° and $\theta_z = 90$ °, corresponding to $\varphi = 90$ ° and $\varphi = 270$ ° from Eq. (2). For this reason, CombR cannot be observed in data shown in Fig. 3(c) around these two angles.

**Discussion**

Our measurements determine precise low energy band parameters for Cd$_3$As$_2$. We demonstrate that $\delta > 0$, implying the presence of two band crossings near the Γ-point. Our value of $\delta$ agrees with the upper bound determined by previous magnetooptical measurements, [6,29] and with the value found in recent density functional theory calculation. [32] With $\delta$ small, and with $E_g$ also small, we expect the two band crossing points to be separated by $2k_0 = 0.012$ Å$^{-1}$ at $B = 0$. The heavy-hole mass is not known, but its impact of the Weyl nodes is minimal, given the very light conduction electrons (see appendix B). However, we believe that additional studies are required to elucidate the impact of the magnetic field on promoting or suppressing possible Weyl nodes. It is also important to mention that strain may have a significant influence on the value of $\delta$. However, from previous studies on the MBE growth of Cd$_3$As$_2$ on GaSb, [34] [35] we can conclude that the thick epilayers studied in this work are relaxed to the bulk crystal structure.

| $E_g$ [meV] | $P_\perp$ [eV·Å] | $\eta^2$ | $\delta$ [meV] | $\Delta_{SO}$ [meV] |
|---|---|---|---|---|
| - 95±5 | 7.96±0.25 | 1.20±0.02 | 15±5 | 400 |

**Table I.** Band parameters for Cd$_3$As$_2$ obtained in this work using a Kane model. $\Delta_{SO}$ is fixed at 400 meV.

In conclusion, we have used magnetooptical spectroscopy to study the band structure parameters of Cd$_3$As$_2$, and have succeeded in establishing that $\delta = 15 \pm 5$ meV. The band parameters extracted from this study are shown in Table I. The behavior of the band structure of this material under a magnetic field has been a subject of considerable debate. Recent works have argued that at high magnetic field, carriers in Cd$_3$As$_2$ are perfectly described by the Kane model, and only for $E_F \sim \delta$ one can observe Dirac or Weyl fermions. [6,29] Our measurements corroborate this result but also establish the necessary presence of a non-zero crystal field splitting in Cd$_3$As$_2$ whose presence yields Weyl fermions under a magnetic field.

Additional measurements will be required to investigate the exact magnetic field range required for Weyl fermions to arise and the exact magnetic field at which they might annihilate. Those measurements will be of great interest to understand the behavior of the quantum Hall effect recently observed in $Cd_3As_2$ quantum wells. [35 - 38]

**Acknowledgements**: Work at ENS is partially supported by ANR-LabEx grant ENS-ICFP ANR-10-LABX-0010/ANR-10-IDEX-0001-02PSL. Work at UCSB was supported through the Vannevar Bush Faculty Fellowship program by the U.S. Department of Defense (grant no. N00014-16-1-2814). T.S. and S.S. thank Luca Galletti and David Kealhofer for discussions. B.A.A. and G.K. acknowledge discussions with B. Bradlyn and J. Cano about the Kane model.

### APPENDIX A: Description of the tetragonal $Cd_3As_2$ band structure by a modified 8-band Kane model

The $Cd_3As_2$ band structure near the Γ point can be described by a $\mathbf{k}.\mathbf{p}$ model which extends the Kane model for cubic symmetry with an additional tetragonal crystal field splitting $\delta$ along the [001] direction. [25,26] The rotational symmetry along this axis prevents the band hybridization in this direction and allows the existence of Dirac nodes if two bands cross. The band structure in the vicinity of $\mathbf{k} = 0$ is described by the following Hamiltonian: [6]

$$H(k) = \begin{pmatrix} E_g & P_\perp k_- & -P_\perp k_+ & 0 & 0 & 0 & 0 & P_\| k_z \\ P_\perp k_+ & 0 & 0 & 0 & 0 & 0 & 0 & 0 \\ -P_\perp k_- & 0 & -2\delta/3 & \sqrt{2}\Delta_{SO}/3 & 0 & 0 & 0 & 0 \\ 0 & 0 & \sqrt{2}\Delta_{SO}/3 & -(\delta+\Delta_{SO}/3) & P_\| k_z & 0 & 0 & 0 \\ 0 & 0 & 0 & P_\| k_z & E_g & P_\perp k_+ & P_\perp k_- & 0 \\ 0 & 0 & 0 & 0 & P_\perp k_- & 0 & 0 & 0 \\ 0 & 0 & 0 & 0 & P_\perp k_+ & 0 & -2\delta/3 & \sqrt{2}\Delta_{SO}/3 \\ P_\| k_z & 0 & 0 & 0 & 0 & 0 & \sqrt{2}\Delta_{SO}/3 & -(\delta+\Delta_{SO}/3) \end{pmatrix}$$

in the basis $i|S\downarrow>, |(X-iY)\downarrow>/\sqrt{2}, -|(X+iY)\downarrow>/\sqrt{2}, |Z\uparrow>, i|S\uparrow>, |(X+iY)\uparrow>/\sqrt{2}, |(X-iY)\uparrow>/\sqrt{2}, |Z\downarrow>$ with the spin and $k_z$ along the [001] direction. The five model parameters are $E_g$, $P_\|$, $P_\perp$, $\Delta_{SO}$, and $\delta$.

In this model, the light electron and hole bands are separated by the energy $E_g < 0$. The crystal field splitting $\delta > 0$ shifts the conduction band down in energy with respect to the heavy hole band yielding two stable crossing points along $k_z$. The heavy hole band is flat. This approximation holds for low-order Landau levels and energies close to the Γ point. The anisotropy of the constant energy surface can be simply evaluated by the factor $\eta^2 = P_\perp/P_\|$.

The resulting calculated band structure is quite similar to those obtained from DFT calculations when comparable band parameters are used. [12]

### APPENDIX B: Remote band effect and the heavy hole effective mass in $Cd_3As_2$

In the $\mathbf{k}.\mathbf{p}$ model developed above, the heavy hole band is flat. It acquires a negative curvature by taking into account the far-band effect following the method developed by Luttinger and Kohn. [40] Similarly to Pidgeon and Brown [41] for narrow-gap semiconductors (InSb, HgTe, …), we modify the Hamiltonian $H(k)$ by introducing the parameters $\gamma_1$, $\gamma_2$, $\gamma_3$ which represent the far-band interactions treated to $k^2$ order.

We use the axial approximation: $\gamma_2 = \gamma_3 = \gamma$. The heavy hole band is then parabolic, and its effective mass is given by:

$$\widetilde{m}_{HH} = \frac{m_0}{\gamma_1 - 2\gamma}$$

As an example, the resulting dispersion with $\widetilde{m}_{HH} = 0.1 m_0$ ($\gamma_1 = 2$ and $\gamma = -4$) is displayed in Fig. 5. All in all, the impact of the heavy hole effective mass on the position of the Dirac nodes and their energy scale is negligible when $\delta$ is small.

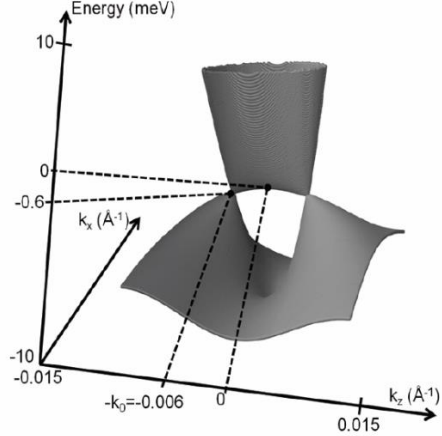

**FIG 5.** Band dispersion of Cd$_3$As$_2$ in the vicinity of the Dirac cones with a heavy hole effective mass fixed at $0.1 m_0$.

### APPENDIX C: $\delta$-dependence of the Fermi surface anisotropy

By solving the Hamiltonian $H(k)$, the constant energy ellipsoidal surfaces in the momentum space take the following form: [20,29]

$$1 = \frac{k_x^2 + k_y^2}{a^2} + \frac{k_z^2}{b^2}$$

where $a^2 = \frac{\gamma(E)P^2}{P_\perp^2 f_{1\theta_z}(E)}$ and $b^2 = \frac{\gamma(E)P^2}{P_\parallel^2 f_{2\theta_z}(E)}$ are the two main axis of the ellipsoid. Therefore, the anisotropy factor can be written as:

$$K = \frac{b}{a} = \eta^2 \sqrt{1 + \frac{\delta(E + \Delta_{SO}/3)}{E(E + 2\Delta_{SO}/3)}} \qquad (3)$$

It is then clear than the anisotropy of the Fermi surface depends on both $\eta^2$ and $\delta$. Figure 6 shows the variation of $K/\eta^2$ with $\delta$ for different Fermi energies. For $E_F = 70$ meV, the $\delta$-dependence of the Fermi surface anisotropy is still significant. Its contribution to the band dispersion of the Landau levels is also enhanced by spin-orbit corrections (last term in Eq. (1)). This makes the determination of $\delta$ possible via the angular dependent measurements.

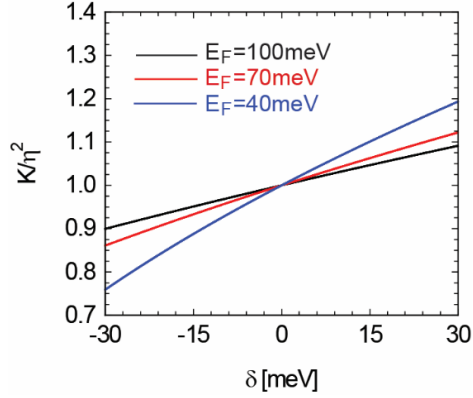

**FIG 6.** Theoretical variation of $K/\eta^2$ with $\delta$ for three different Fermi energies. Curves are calculated using Eq. (3).

**APPENDIX D: Inter-spin coupling terms and observation of the combined resonance**

It is possible to rotate the entire Hamiltonian $H(k)$ into a new coordinate system, in which the $z$ axis coincides with the direction of the magnetic field. [29] This can be done by changing the momentum coordinates through the following transformation:

$$\begin{pmatrix} q_x \\ q_y \\ q_z \end{pmatrix} = \begin{pmatrix} \frac{P_\perp}{P}\cos(\alpha) & 0 & -\frac{P_\parallel}{P}\sin(\alpha) \\ 0 & \frac{P_\perp}{P} & 0 \\ \frac{P_\perp}{P}\sin(\alpha) & 0 & \frac{P_\parallel}{P}\cos(\alpha) \end{pmatrix} \begin{pmatrix} k_x \\ k_y \\ k_z \end{pmatrix}$$

which implies for the new commutation relations:

$$\begin{cases} [q_x, q_y] = -\dfrac{ieB}{h} \\ [q_x, q_z] = 0 \\ [q_y, q_z] = 0 \end{cases}$$

The corresponding Hamiltonian for $q_z // B$ writes:

$$\begin{pmatrix} E_g & Pq_- & -Pq_+ & 0 & 0 & 0 & 0 & Pq_z \\ Pq_+ & -\frac{\delta\sin(\alpha)^2}{2} & \frac{\delta\sin(\alpha)^2}{2} & 0 & 0 & 0 & 0 & \frac{\delta\sin(2\alpha)}{2\sqrt{2}} \\ -Pq_- & \frac{\delta\sin(\alpha)^2}{2} & -\frac{\delta\sin(\alpha)^2}{2} - 2\delta/3 & \sqrt{2}\Delta_{SO}/3 & 0 & 0 & 0 & -\frac{\delta\sin(2\alpha)}{2\sqrt{2}} \\ 0 & 0 & \sqrt{2}\Delta_{SO}/3 & -(\delta\cos(\alpha)^2 + \Delta_{SO}/3) & Pq_z & \frac{\delta\sin(2\alpha)}{2\sqrt{2}} & \frac{\delta\sin(2\alpha)}{2\sqrt{2}} & 0 \\ 0 & 0 & 0 & Pq_z & E_g & Pq_+ & Pq_- & 0 \\ 0 & 0 & 0 & \frac{\delta\sin(2\alpha)}{2\sqrt{2}} & Pq_- & -\frac{\delta\sin(\alpha)^2}{2} & -\frac{\delta\sin(\alpha)^2}{2} & 0 \\ 0 & 0 & 0 & \frac{\delta\sin(2\alpha)}{2\sqrt{2}} & Pq_+ & -\frac{\delta\sin(\alpha)^2}{2} & -\frac{\delta\sin(\alpha)^2}{2} - 2\delta/3 & \sqrt{2}\Delta_{SO}/3 \\ Pq_z & \frac{\delta\sin(2\alpha)}{2\sqrt{2}} & -\frac{\delta\sin(2\alpha)}{2\sqrt{2}} & 0 & 0 & 0 & \sqrt{2}\Delta_{SO}/3 & -(\delta\cos(\alpha)^2 + \Delta_{SO}/3) \end{pmatrix}$$

Off diagonal terms proportional to $\sin(2\alpha) = \sin[2\arctan(\eta^2\tan\theta_z)]$ appear as a result of the axis rotation. These terms yield a spin-orbital mixing of the Landau levels that activates the combined resonance referred to as CombR. The intensity of the CombR is proportional to $|\sin(2\alpha)|^2$ and as a consequence, the CombR is expected to vanish near $\varphi = 90°$ and $\varphi = 270°$ in good agreement with the experimental data shown in Fig. 3(c).